\documentstyle[preprint,aps,eqsecnum]{revtex}

\newcommand{\be}{\begin{equation}}
\newcommand{\ee}{\end{equation}}
\newcommand{\bea}{\begin{eqnarray}}
\newcommand{\eea}{\end{eqnarray}}
\newcommand{\nn}{\nonumber \\}
\newcommand{\p}{\partial}
\def\S{\cal S}
\begin{document}

\tightenlines

\preprint{\small CGPG-97/12-3}

\title{Symmetry reduced Einstein gravity \\
 and generalized $\sigma$ and chiral models}
\author{Abhay Ashtekar$^\dagger$ and Viqar Husain$^\dagger$
\footnote{Present address.}} 

\address{$^\dagger$Center for Gravitational Physics and Geometry,\\ 
Department of Physics, Pennsylvania State University, \\
 University Park, PA 16802-6300, USA\\
and\\
$^*$Department of Physics and Astronomy,\\
University of British Columbia,\\
6224, Agricultural Road\\
Vancouver, BC V6H1B2, Canada.\\}
\bigskip

\maketitle

\begin{abstract}  

Certain features associated with the symmetry reduction of the vacuum
Einstein equations by two commuting, space-like Killing vector fields
are studied. In particular, the discussion encompasses the equations
for the Gowdy $T^3$ cosmology and cylindrical gravitational waves. 
We first point out a relation between the $SL(2,R)$ (or
$SO(3)$) $\sigma$ and principal chiral models, and then show that the
reduced Einstein equations can be obtained from a dimensional
reduction of the standard $SL(2,R)$ $\sigma$--model in {\it three}
dimensions. The reduced equations can also be derived from the action
of a `generalized' {\it two} dimensional $SL(2,R)$ $\sigma$--model
with a time dependent constraint. We give a Hamiltonian formulation of
this action, and show that the Hamiltonian evolution equations for
certain phase space variables are those of a certain generalization of
the principal chiral model. Using these Hamiltonian equations, we give
a prescription for obtaining an infinite set of constants of motion
explicitly as functionals of the space-time metric variables.
 
\end{abstract}

\bigskip



\section{Introduction} 


A central mathematical problem in non-linear field theories of
physical interest is that of gaining `control' over the space of
solutions. One strategy is to map the theory of interest in to
`standard' models which have been well-analyzed. Another is to find
the `full set' of conserved quantities associated with field
equations. Both these strategies have proved to be useful in
two-dimensional field theories, where a number of integrable models
have been known to exist for some time \cite{fadtak,das}.  Now, in
presence of two commuting, space-like Killing vector fields, the four
dimensional, vacuum Einstein theory reduces to a two dimensional
field theory.  Hence, it is natural to explore the structure of this
sector of general relativity using techniques that have been
successful in standard integrable models.

If the Killing fields under consideration are hypersurface orthogonal,
the theory has only one local degree of freedom, and the field
equations can in effect be reduced to a rather simple linear partial
differential equation. The model is then exactly soluble and one can
explicitly write down the full set of conserved quantities
\cite{berger,torre,beetle}. On the other hand, when the Killing fields
fail to be hypersurface orthogonal, the theory is both more difficult
and more interesting: Now, there are {\it two} local degrees of
freedom which remain coupled non-linearly. The space-times in this
category include the Gowdy cosmological models \cite{gow} and
cylindrical gravitational waves (see \cite{kuchar} and references
therein). In this paper, we shall focus on these models.

Various aspects of the underlying equations have been studied in
detail over the years (see, e.g.,
\cite{geroch,kinnchit,maison,belzak,hausern,wu,french,kn}).  The
results have led to a wealth of insights in to the structure of the
space of solutions of the reduced theory which, in turn, have made
these `midi-superspaces' convenient tools to probe physical issues
that arise in the full theory, such as quantization and cosmic
censorship.  Quantization of the one polarization cylindrical waves
was studied in \cite{kuchar} and more recently, in \cite{am}. The
second analysis was subsequently used to show that, in this model,
there exist surprisingly large quantum gravity effects which undermine
the validity of the classical and semi-classical theories in
unexpected ways \cite{aa}. More recently, a quantization of the more
interesting two-polarization gravitational waves has also been
achieved \cite{ks2} and its physical implications are now being
explored.  Finally, the models have proved to be useful in analyzing
cosmic censorship in cosmological \cite{vince} as well as
asymptotically flat contexts \cite{bcm}.

The investigations of
\cite{geroch,kinnchit,maison,belzak,hausern,wu,french,kn} have also
shed light on the mathematical structure of the reduced Einstein
equations.  In particular, the presence of the infinite dimensional
Geroch group which acts transitively on the space of solutions
indicates that the system is in some sense exactly integrable. The
application of the inverse scattering methods has also led to results
supporting this view point.  More recently there has been work on
obtaining constants of motion for the reduced Einstein equations with
open spatial topology in \cite{ks1} and closed spatial topology in
\cite{vh}.  However, it is fair to say that we do not yet know if
there exists a {\it complete} set of constants of motion; unlike in
the standard, two-dimensional integrable models \cite{fadtak,das}, an
analysis involving, e.g., bi-symplectic structures has not been
carried out.  Hence it is of interest to seek new avenues to this
problem.

The purpose of this paper is two fold: i) to point out new relations
between the reduced Einstein system and the other well-known models
which might be useful in future investigations; and, ii) to propose a
new strategy to address the problem of finding explicit constants of
motion.  More precisely, we will study the reduced Einstein equations
in the metric variables, further develop the interplay between these
equations and the standard and generalized $\sigma$--models in three
and two dimensions respectively, and, using this relation, provide a
new avenue for obtaining an infinite set of constants of motion.

The plan of the paper is as follows. In Sec. II we recall the reduced
Einstein equations for the Gowdy $T^3$ cosmology and cylindrical
gravitational waves. In Sec. III, we first describe a relation between
the $SL(2,R)$ or $SO(3)$ $\sigma$--models and the principal chiral
model in any space-time dimension, and then show that the reduced
Einstein system is a sub-system of the standard, {\it three}
dimensional $SL(2,R)$ $\sigma$--model. In Sec. IV we elaborate on
the relation between a generalized $\sigma$--model in {\it two}
dimensions, and present a Hamiltonian formulation and associated
evolution equations. This framework is utilized in Sec. V to give a
prescription for obtaining an infinite set of constants of motion.


\section{Preliminaries: Reduced Einstein equations}


Recall that solutions to the vacuum Einstein equations with two
commuting, space-like, Killing vectors fall into classes of well
known space-times. These include plane, cylindrical \cite{kuchar} and
`toroidal' \cite{schmidt} gravitational waves for non--compact spatial
topology and the $T^3$, $S^1\times S^2$ and $S^3$ Gowdy cosmological
models for compact spatial topology \cite{gow}. For definiteness, we
will consider one representative from each class, namely the $T^3$
cosmological model and cylindrical gravitational waves. Although this
material is well-known, it is included here because the final results
will serve as points of departure for our main analysis.

\subsection{Two polarization Gowdy model}

The $T^3$ Gowdy model has the space-time metric 
\be
ds^2 = e^{2A}\ (-dt^2 + d\theta^2) + g_{ab}\ dx^adx^b\ , 
\label{t3}
\ee
where $A=A(t,\theta)$, $x^a = (x^1,x^2)$, and the $2\times 2$ metric 
\be
g_{ab} = R \left( \matrix{ {\rm cosh}W + {\rm cos}\Phi\ {\rm sinh}W &
                             {\rm sin}\Phi\ {\rm sinh}W \cr
               {\rm sin}\Phi\ {\rm sinh}W & 
               {\rm cosh}W - {\rm cos}\Phi\ {\rm sinh}W } \right)
\label{22}
\ee 
is parametrized by the three functions $R(t,\theta)$,
$\Phi(t,\theta)$ and $W(t,\theta)$. The metric (\ref{t3}) has two
commuting space-like Killing vector fields 
\be \left({\p\over \p x^1}\right) \ \ \ \ {\rm and} 
\ \ \ \ \left({\p\over \p x^2}\right)\ , \ee 
with the additional restriction that each of their orbits has the
topology of a two torus.  The $t,\theta$ coordinates have ranges $0 <
t < \infty$ and $0\le \theta \le 2\pi$. With these conventions, the
metric is that of the 3-torus Gowdy cosmology \cite{gow}.  The vacuum
Einstein equations in the Gowdy gauge, defined by setting
$R(t,\theta)=t$, give the two coupled two-dimensional evolution
equations
\bea \ddot{W} + {1\over t} \dot{W} -
W'' + {\rm sinh}W\ {\rm cosh}W\ ( \Phi'^2 - \dot{\Phi}^2 ) &=& 0\ ,
\label{evo1}\\
\ddot{\Phi} + {1\over t} \dot{\Phi} -\Phi'' 
+ 2\ {{\rm cosh}W\over {\rm sinh}W}\ (\dot{\Phi}\dot{W} -\Phi'W') 
&=& 0\ ,
\label{evo2}
\eea
for  $W(t,\theta)$ and $\Phi(t,\theta)$, and the two `constraint' 
equations
\bea
\dot{A} + {1\over 4t} - {t\over 4}\ 
[\ \dot{W}^2 + W'^2  + {\rm sinh}^2W\  (\dot{\Phi}^2 +\Phi'^2)\ ] 
&=& 0\ ,\label{gc1}\\
A' - {t\over 2}\ (\dot{W} W' + {\rm sinh}^2W\ \dot{\Phi}\Phi' ) 
&=& 0\ \label{gc2}. 
\eea
Note that the evolution equations involve only $W(t,\theta)$ and
$\Phi(t,\theta)$ and, given a solution to these equations, the field
$A(t,\theta)$ can be obtained by a simple integration of the
constraint equations. (Eqns. (\ref{evo1}), (\ref{evo2}) also serve as
the consistency conditions for the two constraints.)  

However, because the spatial topology is compact, there is a
subtlety: We can solve for $A$ consistently only if the integral over
the circle of the expression defining $A'$ in Eqn. (\ref{gc2})
vanishes.
\footnote{We thank the referee for drawing our attention to this 
point.}
This leads to a  global constraint relating the two fields
$W(t,\theta)$ and $\Phi(t,\theta)$, namely,
\be
\int_0^{2\pi} d\theta\ 
(\dot{W} W' + {\rm sinh}^2W\ \dot{\Phi}\Phi' )=0\ \label{globc}, 
\ee
which is preserved by the evolution equations. Thus, to obtain a
solution to the Gowdy model, we need to solve (\ref{globc}) and the
evolution equations (\ref{evo1}) and (\ref{evo2}). However, as we will
see, in the Hamiltonian framework, the global constraint generates
just the rigid rotation around the circle. Therefore, it has rather
simple Poisson brackets with other constraints and constants of
motion. Hence, the heart of the problem lies in the evolution
equations. We shall initially focus on these, and incorporate
the global constraint at the end.

The equations, (\ref{evo1}--\ref{evo2}), may be derived from the
action 
\bea S_G (W,\Phi) &=& {1\over 2} \int dt d\theta\ t\ [\ \dot{W}^2 -
W'^2 + {\rm sinh}^2W (\dot{\Phi}^2 - \Phi'^2)\ ] \nn 
&=& {1\over 2} \int dt d\theta\ t\ \sqrt{-\eta}\
\eta^{ab}G_{AB}(Y)\ \p_a Y^A\p_b Y^B\ ,
\label{SG}
\eea
where $Y^1=W$, $Y^2 =\Phi$, $\eta^{ab}={\rm diag}(-,+)$,
$a,b,\cdots=t,\theta$, and $G_{AB}(Y)\ dY^A dY^B= dW^2 + {\rm
sinh}^2W\ d\Phi^2$ is the unit hyperboloid metric.  The Gowdy model
with one polarization is obtained by setting $\Phi =0$ in the above
equations. This leads to a linear evolution equation for $W$, and
hence exact solvability.

\subsection{Cylindrical gravitational waves} 

The space-time metric for cylindrical gravitational waves can be put
into the general form (\ref{t3}--\ref{22}) above.  The line element is
\be ds^2 = e^{2A}\ (-dt^2 + dr^2) + g_{ab}\ dx^adx^b\ . \ee
The spatial topology is now $R^3$, the orbits of the two Killing
vector fields has topology $S^1\times R$, and all the metric functions
depend only on $t$ and $r$.  Now the relevant gauge fixing is
$R(t,r)=r$. The vacuum Einstein equations are very similar to
(\ref{evo1}--\ref{evo2}); all the $t$ and $r$ factors are interchanged
with appropriate sign changes.  The evolution and constraint equations
are 
\bea \ddot{W} - {1\over r}\ W' - W'' + {\rm sinh}W\ {\rm cosh}W\ (
\Phi'^2 - \dot{\Phi}^2 ) &=& 0\ , \label{cyl1} \\ \ddot{\Phi} -
{1\over r}\ \Phi' - \Phi'' + 2\ {{\rm cosh}W\over {\rm sinh}W}\
(\dot{\Phi}\dot{W} -\Phi'W') &=& 0\ , \label{cyl2} \eea and \bea A' +
{1\over 4r} - {r\over 4}\ [\ \dot{W}^2 + W'^2 + {\rm sinh}^2W\
(\dot{\Phi}^2 +\Phi'^2)\ ] &=& 0\ , \\ \dot{A} - {r\over 2}\ (\dot{W}
W' + {\rm sinh}^2W\ \dot{\Phi}\Phi' ) &=& 0\ , \eea 
where now ${}^\prime \equiv \p/\p r$. Since the spatial topology is
now $R^1$, there is no longer a global constraint. We can therefore
focus simply on the evolution equations. These follow from the action
\be S_{C}(W,\Phi) = {1\over 2} \int dt dr\ r\ [\ \dot{W}^2 - W'^2 +
{\rm sinh}^2W (\dot{\Phi}^2 - \Phi'^2)\ ]\ .
\label{SC}
\ee
The only difference in the Lagrangian densities of (\ref{SG}) and
(\ref{SC}) is the interchange of the $t$ and $r$ factors.

The awkward feature of these actions is the explicit factor of $t$ or
$r$ in the respective integrands. Without these factors, the actions
would be those of the usual two-dimensional non--linear
$\sigma$--model, and standard methods for obtaining complete sets of
conserved currents would be applicable. In the present case, these
procedures have to be modified.

We will discuss two possible avenues.  The first is to embed the two
dimensional model (\ref{SG}) in a standard three dimensional, flat space
non-linear $\sigma$-model, which is free of the $t$ (or $r$)
factor. (The two-dimensional model, with its $t$ (or $r$) factor is
recovered as a dimensional reduction of the standard three dimensional
model.) In this approach, the solutions of the reduced Einstein theory
constitute a specific (symmetry-reduced) sector of the solution space
of the three dimensional $\sigma$-model. The second avenue is to remain 
in two dimensions and handle the $t$ (or $r$) factor by imposing a $t$
(or $r$) dependent constraint on a free theory of three scalar
fields. We will explore these avenues in sections III and IV
respectively.


\section{Relation to three  dimensional $\sigma$ and chiral models}


In Section III.A, we point out a relation between $SL(2,R)$ or
$SO(3)$ $\sigma$-models and principal chiral models for these groups
which exists in {\it any} space-time dimension but which appears not to
have been noted in the literature. In Section III.B, we return to the
reduced Einstein system and show that its space of solutions can be
naturally embedded in to the space of solutions of the $SL(2,R)$
$\sigma$--model in three space-time dimensions. Thus, the three models
are closely related; the reduced Einstein theory is a sub-case of the
other two models. In particular, any method of showing integrability
or extracting conserved quantities in three dimensional $\sigma$ or
chiral models is directly applicable to the reduced Einstein system.

\subsection{$\sigma$--model and flat connections}

For concreteness let us consider the $SL(2,R)$ non-linear
$\sigma$-model in $n$-dimensional Minkowski space-time; with obvious
sign changes, our discussion goes through step by step in the
$SO(3)$ case and for Euclidean space-time signature. The model has the
action
\be S_n (X,\lambda) =  -{1\over 2}\, 
\int d^nx\ \sqrt{-\eta}\ \left[\ \eta^{ab}\p_aX^i\p_bX^j g_{ij} +
\lambda\ ( g_{ij}X^iX^j + 1 )\ \right]\ ,
\label{sn}
\ee
where $\eta^{ab}$ is the flat Lorentzian metric of signature
$(-,+,\cdots,+)$, $g_{ij}$ is the Cartan metric for $SL(2,R)$ of
signature ${\rm diag}(++-)$, and $\lambda(x)$ is a Lagrange
multiplier. The equations of motion are:
\bea
\Box X^i \equiv \eta^{ab}\p_a\p_b X^i &=& \lambda X^i
\label{sigmaeom}\\
g_{ij}X^iX^j + 1 &=& 0\, . \label{conn}
\eea

To see the relation to the principal chiral model, define a Lie
algebra valued 1--form by
\be
A^i_a = -2 f^i_{\ jk} X^j\p_aX^k,
\ee
where $f^i_{\ jk}$ are the $SL(2,R)$ structure constants. Then it 
follows from the equation of motion (\ref{sigmaeom}) that
\be 
\p^a A_a^i = 0\ . 
\label{pch1}
\ee
Furthermore, using $X^i\p_aX_i = 0$, which follows from the 
constraint (\ref{conn}), it is easy to show that 
\be 
F_{ab}^i = \p_a A_b^i -\p_b A_a^i + f^i_{\ jk}A_a^jA_b^k = 0\ . 
\label{pch2}
\ee
Thus, the $\sigma$--model equations imply (\ref{pch1}) and
(\ref{pch2}), which are the equations of the principal chiral model.
The last step makes a crucial use of the three dimensionality of our
Lie groups (i.e., the specific form of the structure constants of
$SL(2,R)$ (or $SO(3)$)). Hence, while this relation holds in {\it any}
space-time dimension, it does {\it not} hold for general groups.

Let us now ask if this map from the the space of solutions of the
non-linear sigma model to that of the principal chiral model is
surjective. Can we derive (\ref{sigmaeom}--\ref{conn}) from the chiral
model equations? First, $F_{ab}=0$ implies that $A_a$ has the form
$A_a = g^{-1}\p_a g$ for $g\ \epsilon\ SL(2,R)$. Now, a standard
parametrization of $g$ is
\be 
 g = \left( \matrix{ \alpha + X^2 & X^1 + X^3 \cr
                     X^1 - X^3 & \alpha - X^2 }
            \right)\ , 
\label{matrix}
\ee
with the unit determinant condition $\alpha^2 - g_{ij}X^iX^j = 1$. 
This gives 
\be 
A_a =  g^{-1}\p_a  g = 2\ (\alpha\p_a {\cal X} - {\cal X}\p_a\alpha) 
                       -2\ [{\cal X}, \p_a{\cal X}]\ ,
\ee
where ${\cal X} = X^i\tau_i$ is an element of $sl(2,R)$ with
generators $\tau_i$. Secondly, $\p^a A_a=0$ leads to the equation
\be 
 [{\cal X},\Box {\cal X}] = \alpha \Box {\cal X} 
- {\cal X}\Box\alpha \ .
\ee  
The unit determinant condition implies that the $\sigma$-model
constraint (\ref{conn}) is satisfied if and only if $\alpha =0$.  In
this case, the commutator $ [{\cal X},\Box {\cal X}]$ vanishes which
in turn implies that $\Box X^i = \lambda X^i$ for some function
$\lambda(x)$. Thus, the $\sigma$-model equations hold if and only if
the $SL(2,R)$ matrices $g$ are trace free. To summarize, the space of
solutions to the $\sigma$-model is naturally embedded in the space
of solutions to the principal chiral model. The image consists of flat
connections $A_a$ given by $A_a = g^{-1}\p_a g$ where $g$ is
trace-free (i.e., where $\alpha =0$ in the parametrization
(\ref{matrix})). We will return to this result in the next subsection.

To conclude this discussion, we note that there is also a Hamiltonian
version of the derivation of the $SL(2,R)$ principal chiral model from
the corresponding $\sigma$--model, which is presented below for a more
general time dependent constraint $X^iX_i + t = 0$. As we will see,
this is the type of constraint relevant for the reduced Einstein
equations if one wants to work with two dimensional models.


\subsection{Reduced theory from three dimensions}


Let us now specialize to three-dimensional space-times with topology
$S^1\times R^2$, equipped with a flat metric $\eta_{ab}$, given by
\be
\eta_{ab}dx^adx^b = -d\tau^2 + dx^2 + d\theta^2.  
\ee
Consider the action  
\be S'_3[Y] = -{1\over 2}\, \int_M d\tau dx d\theta\ \sqrt{-\eta}\
\eta^{ab}G_{AB}(Y)\ \p_a Y^A\p_b Y^B,
\label{s3}
\ee
where, as before $A =1,2$ but now $Y^A = Y^A(\tau,x,\theta)$. As is
well-known, this action yields the same equations of motion as
$S_3(X,\lambda)$ of (\ref{sn}). Note, however, that, in contrast with
the reduced Einstein action (\ref{SG}), there is no factor of $\tau$
in the integrand of (\ref{s3}). However, we will now show that (\ref{SG})
does result from (\ref{s3}) by a symmetry reduction. Thus, the reduced
Einstein model is contained in the three dimensional, standard, non-linear
$\sigma$-model.

To perform the required reduction to two dimensions, we proceed in two
steps. First,let us make the change of coordinates
\be
\tau = t\ {\rm cosh} y,\ \ \ \ \ \ x = t\ {\rm sinh} y  
\ee
in the $\tau-x$ plane, which casts the metric in the form
\be
\eta_{ab}\ dx^adx^b = -dt^2 + t^2 dy^2 + d\theta^2\, . 
\ee
Second, let us require that the Lie derivative of the field variables
$Y^A(t,y,\theta)$ along the boost-Killing vector field $(\p/\p y)^a$
of $\eta^{ab}$ vanish. (In the chosen coordinates this means $Y^A =
Y^A(t,\theta)$.)  Then, we have:
\be S'_3[Y] \longrightarrow S'_2[Y] = - {1\over 2}\, \int dy \int dt
d\theta\ t\ \eta^{ab}G_{AB}(Y)\ \p_a Y^A\p_b Y^B = S_G[Y] \left( \int
dy\right)\ , \ee
which is the Einstein action (\ref{SG}) for Gowdy models, multiplied
by the constant $\int dy$.

Thus, we arrive at the following conclusions. As we have just shown,
the space of solutions ${\S}_{G}$ to the reduced Einstein equations
(i.e. to the Gowdy model) is naturally embedded {\it in} the space
${\S}_{\sigma}$ of solutions to the $\sigma$-model resulting from the
action $S'_3[Y]$ of (\ref{s3}), or $S_3[X,\lambda]$ of (\ref{sn}).
Second, we saw in section III.A that the space ${\S}_{\sigma}$ is in
turn embedded into the space of solutions ${\S}_{ch}$ to the principal
chiral model. The image of the first map consists of fields
$Y^A(t,\theta)$ which are $y$-independent, and satisfy the global
constraint (\ref{globc}), while the image of the second map consists
of $A_a = g^{-1}\p_a g$ for which the $g$ of (\ref{matrix}) is
trace-free. Therefore, any method for finding conserved quantities for
the standard non-linear $\sigma$-model or the principal chiral model
in three dimensions yields, via restriction, a method for finding
conserved quantities for the reduced Einstein equations. However, in
practice, this simplification is not directly useful since the
standard techniques for finding conserved currents for $\sigma$ and
chiral models are tailored to two space-time dimensions.  Nonetheless,
the existence of embeddings is conceptually interesting and may have a
more general application. For example, results on integrability,
asymptotic forms of solutions, and quantization of the standard three
dimensional $\sigma$ and chiral models will also apply to Gowdy models
and cylindrical waves by restriction.  Indeed, this is essentially
the procedure used to (find the complete set of constants of motion
and) quantize one polarization cylindrical \cite{kuchar,am},
toroidal \cite{beetle} waves and Gowdy models \cite{monica}.

 
\section{Relation to  two  dimensional generalized $\sigma$ 
and chiral models} 


Let us now explore the relation between the reduced Einstein system
and {\it two} dimensional models. In three space-time dimensions, we
could restrict ourselves to the standard $\sigma$ and chiral
models. In two dimensions, on the other hand, we will have to allow
certain generalizations.

More precisely, in Sec. IV.A, we will show that the reduced Einstein
equations can be derived from a two dimensional generalized
$\sigma$--model action, where, however, the constraint is time (or
space) {\it dependent}. In Sec. IV.B, we will perform a Legendre
transform and show that the Hamiltonian evolution equations for
certain variables give a generalized chiral model. This latter result
is a Hamiltonian version of that given in Sec. III.A above, suitably
generalized to incorporate the space or time dependent $\sigma$--model
constraint.

\subsection{Time dependent $\sigma$--model}

Let the space-time topology be $S^1\times R$ and let $\eta_{ab}$
be the obvious flat metric with signature $(-,+)$.  Consider the
following two-dimensional action for three free scalar fields $X^i$
$(i=1,2,3)$:
\be 
S [X,\lambda] =  -{1\over 2}\, \int dt d\theta\left[ \sqrt{-\eta}\ 
\eta^{ab} 
g_{ij}\ \p_a X^i\p_b X^j - \lambda\ (g_{ij}X^iX^j + t) \right]\ ,
\label{slam}
\ee
where $g_{ij}=$diag$(++-)$. The variation of $S$ with respect to 
$\lambda$ gives the {\it time dependent} 
constraint  
\be
C_1 := g_{ij}X^iX^j + t \equiv \bar X\cdot\bar X + t = 0\ .
\label{cons}
\ee
This constraint is solved by  setting 
\bea
X^1 = \sqrt{t}\ {\rm cos}\Phi\ {\rm sinh} W, \ \ 
X^2 = \sqrt{t}\ {\rm sin}\Phi\ {\rm sinh} W, \ \ 
X^3 = \sqrt{t}\ {\rm cosh} W \, , 
\label{solX}
\eea 
where the fields $W(t,\theta)$ and $\Phi(t,\theta)$ are unrestricted.
These conditions, when substituted in to the action (\ref{slam}), give
the reduced action 
\be S_R[W,\Phi] = {1\over 2} \int dt d\theta\
\left\{\ t\ [\ \dot{W}^2 - W'^2 + {\rm sinh}^2W (\dot{\Phi}^2 -
\Phi'^2)\ ] - {1\over 4t}\ \right\}.
\label{ract}
\ee
The integrand here differs from that of $S_G$ of (\ref{SG}) only by an
additive field independent term, and therefore leads to the same
evolution equations as (\ref{SG}). It is also straightforward to
verify directly that, as is expected on general grounds, the equations
of motion following from $S[X,\lambda]$, together with (\ref{solX}),
lead to the reduced Einstein equations (\ref{evo1}--\ref{evo2}).

\subsection{Hamiltonian formulation and generalized chiral model}

We will now show that the Legendre transform of the action
(\ref{slam}) leads to ``chiral model--like'' equations, where however
the curvature is not flat. The phase space form of the action is:
\be 
S[X,\lambda,P] = \int dt d\theta \left\{\ P_i\dot{X}^i - 
 {1\over 2}\ \left[ g^{ij}P_iP_j 
+ g^{ij}X_i^\prime X_j^\prime - \lambda\ (g^{ij}X_iX_j + t )\ 
\right] \right\}\ .
\label{1+1L}
\ee
where $P_i$ are the momenta conjugate to $X^i$.
Following now the Dirac prescription, we require that the constraint 
$C_1$ of (\ref{cons}) be preserved in time:
\be 
{dC_1\over dt} = \{C_1,H\} + {\p C_1\over \p t} 
= 2\bar X\cdot\bar P + 1 = 0\ , \label{cons2}
\ee
where 
\bea 
H &=& \frac{1}{2} \int_0^{2\pi} d\theta\ g^{ij}(P_iP_j+X'_i X'_j)  
+ {1\over 2}\int_0^{2\pi}d\theta\ \lambda C_1 \nonumber\\
&\equiv& H_0 + {1\over 2}\int_0^{2\pi}d\theta\ \lambda C_1 
\eea 
is the Hamiltonian identified from (\ref{1+1L}).  This gives the
secondary constraint 
\be 
C_2 := 2\bar X\cdot\bar P+1=0.
\ee
The pair of
constraints $C_1=0$ and $C_2=0$ are of second class. The constraint
$C_2$ is solved by setting
\bea
P_1 &=& {1\over 2\sqrt{t}}\ {\rm cos} \Phi\ {\rm sinh} W - 
\sqrt{t}\ {\rm sin} \Phi\ {\rm sinh} W\ \dot{\Phi} + \sqrt{t}\  
{\rm cos}\Phi\ {\rm cosh}W\ \dot{W}\nn 
P_2 &=&  {1\over 2\sqrt{t}}\ {\rm sin} \Phi\ {\rm sinh} W +
\sqrt{t}\ {\rm cos} \Phi\ {\rm sinh} W\ \dot{\Phi} + \sqrt{t}\  
{\rm sin}\Phi\ {\rm cosh}W\ \dot{W}\nn 
P_3 &=& - {1\over 2\sqrt{t}}\ {\rm cosh} W - \sqrt{t}\ {\rm sinh}W\  
\dot{W} 
\label{solP}
\eea
and using (\ref{solX}). As we did for $C_1$, let us require that the
secondary constraint $C_2$ be preserved in time. This will lead either
to a new constraint, or to a condition on the lagrange multiplier
$\lambda$.  It turns out that the latter is the case :
\be 
{dC_2\over dt} = \{C_2, H\} + {\p C_2\over \p t} = 
 \{C_2,H_0\} + 4t \lambda\ = 0\ , 
\ee
which gives $\lambda = -\{C_2,H_0\}/4t$ and 
\be 
H = H_0 - {1\over 8t}\int_0^{2\pi}d\theta\ \{C_2, H_0\}C_1 \ .
\label{ham}\ee
This is the (first class) Hamiltonian we must use to derive evolution
equations. The above procedure is equivalent to using the Dirac
brackets for the unmodified Hamiltonian $H_0$.  To summarize, a
Hamiltonian description of the evolution equations of the Gowdy model
can be given as follows.  The phase space consists of canonical pairs
$(X^i, P_i)$. There are two second class constraints and the
Hamiltonian is given by (\ref{ham}). If we initially satisfy the two
constraints, they continue to be satisfied by the Hamiltonian flow.

Recall however that the equations governing the Gowdy model also
includes a global constraint (\ref{globc}). It is straightforward to
verify that this constraint generates the rigid diffeomorphism on
$S^1$. Hence it weakly Poisson-commutes with $C_1$ and $C_2$ and
strongly Poisson-commutes with the Hamiltonian. Therefore, it is a
single (global) first class constraint within the Hamiltonian
description arrived at above and we can just carry it along in the
analysis. Constants of motion of the system {\it without} the global
constraint will provide physical constants of motion simply by
restriction to the part of the phase space where the global constraint
is satisfied {\it provided} they are invariant under rigid
diffeorphisms of $S^1$.

To go to a generalized chiral model, it is convenient to replace 
$(X^i, P_i)$ pairs by new variables 
\be
L^i = f^i_{\ jk} X^jP^k \ \ \ {\rm and}\ \ \  
J^i= f^i_{\ jk} X^j X^{k\prime},
\ee
where $f^i_{\ jk}$ are the $SL(2,R)$ structure constants. We will now
argue that there is no loss of information in using $L^i, J^i$ in place
of $X^i, P^i$ satisfying the two constraints (\ref{cons}) and
(\ref{cons2}). First, it is easy to verify that both $L^i$ and $J^i$
are space-like in the internal directions (but can vanish). Hence,
$\bar{J}\times \bar{L}$ is time-like. When this last vector is
non-zero, one can easily recover $X^i$ and $P^i$ from $L^i$ and $J^i$
explicitly:
\be 
\bar{X}
= \pm { \sqrt{t}\over \left| \bar{J}\times \bar{L}\right| }
\  \bar{J}\times \bar{L}\ , \ \ \ \ \ \ \ 
\bar{P} = \pm { 1\over \sqrt{t}\left| \bar{J}\times \bar{L}\right| } 
\left( {1\over 2} \bar{J}\times \bar{L} 
+ \left[(\bar{J}\times \bar{L})\times \bar{L}\right] \right)\ ,
\label{explicit}
\ee
where 
\be 
\left| \bar{J}\times \bar{L}\right| 
= \sqrt{-(\bar{J}\times \bar{L})\cdot 
\left(\bar{J}\times \bar{L}\right)}\ .
\ee 
The choice of the sign in the two equations can be fixed by first
requiring that $X^i$ be future-directed and then using (\ref{cons2}).
Since $\bar{J}\times\bar{L}$ is time-like, its norm can vanish only
when the vector itself vanishes. In this case, we can not simply use
explicit formulas (\ref{explicit}). However, a more detailed,
systematic analysis shows that $X^i,P^i$ can be again recovered from
$L^i, J^i$.

$L^i$ and $J^i$ satisfy simple Poisson bracket relations
\bea
    \{ L_i(t,\theta), L_j(t,\theta' \} &=& 
f_{ij}^{\ \ k} L_k(t,\theta)\ \delta(\theta,\theta')\ ,\nn
    \{ J_i(t,\theta), J_j(t,\theta') \} &=& 0\ , \nn 
\{ L_i(t,\theta),J_j(t,\theta') \} &=& 
f_{ij}^{\ \ k}J_k(t,\theta)\ \delta(\theta,\theta') + 
g_{ij}\delta'(\theta,\theta')\ .
\label{pb1}
\eea
Furthermore,
\bea
\{ L^i(t,\theta), C_1(t,\theta') \} &=& 
\{ L^i(t,\theta), C_2(t,\theta') \} = 0 \\
\{ J^i(t,\theta), C_1(t,\theta') \} &=& 0,\ \ \ 
\{ J^i(t,\theta), C_2(t,\theta') \}  
= 4J^i(t,\theta)\ \delta(\theta,\theta')\ . 
\eea
(The indices $i,j,\cdots$ are lowered (raised) using the $SL(2,R)$
metric $g_{ij}$.)  The evolution equations of the $SL(2,R)$ variables 
$(L^i,J^i)$ are:
\be
\dot{L}^i = \{ L^i, H \} = J^{i\prime}, 
\label{ldot}
\ee
where, as before, the prime denotes the derivative with respect to
$\theta$, and
\bea
\dot{J}^i &=& \{ J^i, H_0 \} - \{ J^i, C_1\} \left(1\over 4t\right) 
 \{ C_2, H \}
       = \{ J^i,H_0 \}  \nn
       &= & L^{i\prime} - 2 f^i_{\ jk}X^{j\prime} P^k\ . 
\eea
The last equation can be further simplified by noting that 
\bea 
P^i &\approx& {1\over 2t}\ X^i + {1\over t}\ f^i_{\ jk} X^j L^k ,\\
f^i_{\ jk} X^{j\prime} P^k &\approx& -{J^i\over 2t} 
+{1\over t}\ X^i(\bar X'\cdot\bar L) 
\nn
&\approx& -{J^i\over 2t} + {1\over t}\ f^i_{\ jk} L^j J^k, 
\eea  
where $\approx$ denotes equality modulo the constraints $C_1$ and
$C_2$.  This finally gives
\be
\dot{J}^i = L^{i\prime} - {2\over t} f^i_{\ jk} L^j J^k 
+ {1\over t} J^i. 
\label{jdot}
\ee 

We are now ready to put these Hamiltonian equations in to a chiral 
model--like form. Define the matrices
\be 
 A_0 := 2 L^i\tau_i\ , \ \ \ \ \ \ A_1 := 2 J^i\tau_i\ ,
 \ee
where  
\be 
   \tau_1 = {1\over 2  }\left( \matrix{
         0 & 1 \cr
         1 & 0  } \right)\ , \ \ \
   \tau_2 = {1\over 2  }\left( \matrix{
         1 & 0 \cr
         0 & -1} \right)\ ,\ \ \
   \tau_3 = {1\over 2  }\left( \matrix{ 
         0 & 1 \cr
	 -1 & 0 }  \right)\ ,
\ee
are the generators of $SL(2,R)$ satisfying the relations
\be
\left[\tau_i, \tau_j\right] = f_{ij}^{\ \ k}\tau_k\ ,\ \ \ \ \ 
\tau_i \tau_j= {1\over 2} f_{ij}^{\ \ k}\tau_k\ .
\ee
Multiplying the evolution equations (\ref{ldot}) and (\ref{jdot}) 
by $\tau^i$ leads to
\bea 
\p_0 A_0 - \p_1 A_1 &=& 0 \label{ch1}\\
\p_0 A_1 - \p_1 A_0 +  \left[A_0, A_1\right] &=& 
\left(1 - { 1\over t}\right)\ \left[ A_0, A_1 \right] 
+ {A_1\over t}\ ,
\label{ch2}
\eea
where $0,1=t,\theta$; $t>0$ and $0\le \theta <2\pi$. This is the 
desired chiral model-like form of the Einstein evolution equations 
(\ref{evo1}--\ref{evo2}).

So far, we have considered the Gowdy model equations where the
relevant $\sigma$--model constraint is time dependent. The analysis of
cylindrical waves is completely analogous. The evolution equations
(\ref{cyl1}--\ref{cyl2}) may be derived from an action like
(\ref{slam}), but with $r$ replacing $t$ in the constraint. After
performing a Hamiltonian decomposition, and following steps similar to
the above, we find that the cylindrical wave evolution equations for
the variables $L_i$ and $J_i$ lead to the chiral model--like equations
\bea 
\p_0 A_0 - \p_1 A_1 &=& 0\ , \label{cch1} \\
\p_0 A_1 - \p_1 A_0 +  \left[ A_0, A_1\right] &=& 
\left(1 - { 1\over r}\right)\ \left[ A_0, A_1 \right] 
+ {A_0\over r}\ .
\label{cch2}
\eea
Note that these are different from the Gowdy model equations
(\ref{ch1}--\ref{ch2}) only in the replacement $1/t,\, A_1/t
\rightarrow 1/r,\,A_0/r$. In addition, now, there is no analog of the
global constraint (\ref{globc}) to worry about.


\section{Constants of motion}


For the chiral model in two space-time dimensions, there exist
standard techniques to construct an infinite hierarchy of conserved
currents. In this section, we will show that one of them admits a
suitable extension which is applicable to the generalized chiral model
of Sec. IV.


\subsection{Gowdy cosmology}


The idea is to use the Hamiltonian equations (\ref{ch1}--\ref{ch2}) to
extract constants of motion for the reduced Einstein equations. The
method is a modification of one given by Brezin et. al. (BIZZ) for obtaining
constants of motion of the principal chiral model \cite{bizz}. Let us
first rewrite the evolution equation (\ref{ch2}) in a convenient form:
\be
F_{01} =  A_1 + (1-t) \epsilon^{ab}\p_a A_b\ .
\label{ch22}
\ee
In the standard chiral model, the right side vanishes. The
divergence-free condition (\ref{ch1}) on $A_a$ is the same in the two
cases. So, modification of the (BIZZ) procedure of \cite{bizz} is
necessary to handle the right side of (\ref{ch22}).

As for the principal chiral model, (\ref{ch1}) is already a
conservation equation, so we can write the first conserved current as
\be 
J_a^{(1)} = D_a\lambda^{(0)}\ ,
\ee
where $D_a = \p_a + A_a$ and $\lambda^{(0)}$ is any constant matrix. 
Denote the corresponding conserved charge by $Q^{(1)}$:
\be 
Q^{(1)} = \int_0^{2\pi}d\theta\ J_0^{(1)} = 
\int_0^{2\pi}d\theta\ \left[A_0,\lambda^{(0)}\right]
\ee 
Since $J_a^{(1)} - (Q^{(1)}/2\pi)\p_a t =: J_a^{(1)} - \p_a
\phi^{(1)}$ is a smooth conserved current with zero charge, it is dual
of the gradient of a smooth field.  Thus, there exists a smooth,
matrix-valued field $\lambda^{(1)}$ such that
\be J_a^{(1)} - \p_a \phi^{(1)} = \epsilon_a{}^b \p_b\lambda^{(1)}
\, . \label{j1}\ee
This equation determines $\lambda^{(1)}$ up to an additive constant
matrix:
\be 
\lambda^{(1)}(\theta,t) = -\int_0^\theta d\theta'\ J_0^{(1)}(\theta',t)
                   +  {Q^{(1)}\theta \over 2\pi}
                   - \int_{t_0}^t dt'\ J_1^{(1)}(0,t')\ .  
\label{LL1}
\ee 
Note that, in spite of the explicit appearance of $\theta$ in the
second term, the `potential' $\lambda^{(1)}$ is smooth everywhere,
including the point $\theta =0$. It depends on $A_a$, i.e., on the
solution to the Gowdy model under consideration and the idea is to
construct another conserved current from $\lambda^{(1)}$.
 
As in the (BIZZ) procedure of \cite{bizz}, let us introduce a fiducial
current,
\be 
K_a^{(2)} := D_a \lambda^{(1)}\ ,
\ee
and compute its divergence
\bea 
\eta^{ab}\p_a K_b^{(2)} &=& \eta^{ab}D_b\p_a\lambda^{(1)} 
       = \epsilon^{ab}D_a\left(J_b^{(1)}-\p_b \phi^{(1)}\right)\nn
&=& \epsilon^{ab}D_aD_b \lambda^{(0)} 
        - \epsilon^{ab} \left[A_a, \p_b \phi^{(1)}\right] \nn
&=&  \left[ F_{01},\lambda^{(0)}\right] 
      -  \epsilon^{ab} \left[ A_a, \p_b \phi^{(1)}\right] \nn
&=&   \left[A_1, \lambda^{(0)} \right] 
     + (1-t)\epsilon^{ab}\left[ \p_aA_b, \lambda^{(0)} \right]  
     -  \epsilon^{ab} \left[A_a, \p_b \phi^{(1)}\right] \nn
&=& \epsilon^{ab} \p_a\left[ (1-t)A_b, \lambda^{(0)}\right] 
    +  \epsilon^{ab} \delta_a^0 \left[A_b, \lambda^{(0)}\right] 
    +  \left[ A_1, \lambda^{(0)}\right] 
    - \epsilon^{ab}\left[ A_a, \p_b\phi^{(1)}\right]\ , 
\eea
where the equation of motion (\ref{ch22}) is used in the fourth 
equality. Collecting terms, we obtain
\be 
\eta^{ab}\p_a K_b^{(2)} = \epsilon^{ab}
      \p_a\left[(1-t)A_b,\lambda^{(0)}\right] 
    + 2 \epsilon^{ab} \delta_a^0\left[A_b, \lambda^{(0)}\right] 
    + \epsilon^{ab}\delta_a^0\left[A_b, {Q^{(1)}\over 2\pi}\right]. 
\ee
Now the key observation is that the last two terms on the r.h.s. may 
be rewritten as divergences. Indeed, using (\ref{j1}), we obtain
\be 
\epsilon^{ab} \delta_a^0\ \left[A_b, \lambda^{(0)}\right] 
=  \epsilon^{ab}\delta_a^0\left(\epsilon_b^{\ c}\p_c \lambda^{(1)} 
    + \p_b\phi^{(1)}\right)
= \eta^{ab}\p_a\left(\lambda^{(1)}\delta_b^0 
                    + \phi^{(1)} \delta_b^1\right)\ .
\ee
Similarly, $[A_a, Q^{(1)}/2\pi]$ is a conserved current. Using it in 
place of $J^{(1)}_a = [A_a, \lambda^{(0)}]$ in the steps that led us 
to Eqn. (\ref{LL1}), we obtain
\be 
 \epsilon^{ab}\delta_a^0\ \left[A_b, {Q^{(1)}\over 2\pi}\right] 
= \epsilon^{ab} \delta_a^0 
     \left(\epsilon_b^{\ c}\p_c \bar{\lambda}^{(1)} 
    + \p_b\bar{\phi}^{(1)}\right)
= \eta^{ab}\p_a\left(\bar{\lambda}^{(1)}\delta_b^0 
  + \bar{\phi}^{(1)} \delta_b^1\right)\ . 
\ee 
Here $\bar{\lambda}^{(1)}$ is defined similarly to $\lambda^{(1)}$ 
of Eq. (\ref{LL1}) by 
\be 
\bar{\lambda}^{(1)}(\theta,t) = 
 -\int_0^\theta d\theta'\ \left[A_0(\theta',t),Q^{(1)}/ 2\pi\right]
       +  {\bar{Q}^{(1)}\theta\over 2\pi}
       - \int_{t_0}^t dt'\ \left[A_1(0,t'), {Q^{(1)}\over 2\pi}\right]\ ,  
\label{LLB1}
\ee
with  $\left[A_a,Q^{(1)}/ 2\pi\right]$ replacing
$J_a^{(1)} = \left[A_a,\lambda^{(0)}\right]$, and 
$\bar{\phi}^{(1)} = \bar{Q}^{(1)}t/2\pi$ with 
\be 
\bar{Q}^{(1)} = \int_0^{2\pi} d\theta\ 
\left[A_0(\theta,t), Q^{(1)}/ 2\pi\right]\ .
\ee 

Thus, we have shown that the current 
\be
J_a^{(2)} \equiv  K_a^{(2)} 
- \epsilon_a^{\ b}(1-t)\left[A_b,\lambda^{(0)}\right]
-\delta^0_a\left( 2 \lambda^{(1)} + \bar{\lambda}^{(1)} \right) 
- \delta^1_a \left( 2 \phi^{(1)} + \bar{\phi}^{(1)} \right)
\ee
is conserved. The corresponding conserved charge is 
\be 
Q^{(2)} = \int_0^{2\pi}d\theta\ J_0^{(2)}(\theta,t) 
 = \int_0^{2\pi}d\theta\ \left\{\ D_0\lambda^{(1)} - 
\left( 2\lambda^{(1)} + \bar{\lambda}^{(1)} \right) 
 + (1-t)\left[ A_1,\lambda^{(0)} \right]\ \right\}\ , 
\ee 
with $\lambda^{(1)}$ and $\bar{\lambda}^{(1)}$ are defined as in 
(\ref{LL1}) and (\ref{LLB1}). 

Notice that $Q^{(1)}=Q^{(1)}_i\tau_i$ and $Q^{(2)}=Q^{(2)}_i\tau_i$
are independent charges because the latter depends explicitly on $t$,
$A_0$ and $A_1$, whereas the former depends only on $A_0$. These are
in fact six different charges because we can ``peel off'' the matrices
$\tau_i$. By inspection, these charges are invariant under rigid
diffeomorphisms of $S^1$. Hence, they Poisson-commute with the global
constraint (\ref{globc}) and are physical constants of motion for the
Gowdy model.

The key question now is whether we can build a whole tower of
conserved charges. A natural strategy is to use Poisson brackets
between the $Q^{(1)}_i$ and $Q^{(2)}_i$.  The brackets between
$Q^{(1)}_i$ among themselves form the $sl(2,R)$ Lie algebra; they do
not yield new charges. Moreover, their brackets with $Q^{(2)}$ just
rotate the $Q^{(2)}_i$ among themselves. On the other hand, the
Poisson brackets $\{ Q^{(2)}_i, Q^{(2)}_j \}$ provides charges whose
expressions have a higher degree of non-locality than the $Q^{(2)}_i$
themselves; the expressions contain one more nested integral. It is
therefore reasonable to suppose that, generically, the Poisson bracket
is a {\it new} conserved charge.  It is straightforward to see that
the continuation of this process gives conserved charges of
successively higher degrees of non-locality, each of which appear to
be functionally independent of the previous ones.  Thus, the procedure
yields an infinite tower of conserved quantities.  Although we do not
have a conclusive proof, the structure of non-locality suggests that
all these charges are independent. Finally, by construction, all
these charges Poisson-commute with the global constraint (\ref{globc})
and are therefore physical charges for the Gowdy model.
 
Let us summarize. The equations of motion following from the action
(\ref{slam}) are the reduced Einstein equations. We performed a
$1+1$--decomposition of this action to obtain the Hamiltonian theory,
and then made a change of variables from $(X^i,P_i)$ to $A_a^i$. The
equations for $A_a^i$ resemble those of the chiral model. Hence, it is
possible to extend the standard procedure \cite{bizz} to obtain two
sets of conserved charges $Q^{(1)}_i$ and $Q^{(2)}_i$. A tower of new
conserved quantities can be obtained by taking Poisson brackets.
Finally, note that the constants of motion can be rewritten in terms of
the original metric variables $(W,\Phi)$ and their space and time 
derivatives by using the solutions (\ref{solX}) and (\ref{solP}) of
the constraints $C_1$ and $C_2$.


\subsection{Cylindrical waves} 


As noted before, the evolution equations for cylindrical waves and the
$T^3$ Gowdy model are very similar. Therefore it is not surprising
that the above approach to finding constants of motion applies to
cylindrical waves. There are two main differences. First, the spatial
topology is now $R^3$ rather than $T^3$, whence the spatial sections
of the reduced two-dimensional model are half-lines $0\le r$ rather
than circles. The triviality of this topology will simplify the
analysis.  The second difference is that we now need to impose
boundary conditions on our dynamical variables. For simplicity, we
will assume that, on an arbitrarily chosen initial Cauchy surface, the
fields $W$ and $\Phi$ are of compact support. This assumption will
ensure the convergence of various integrals but can be weakened in an
obvious manner.

As in the Gowdy model, the existence of the first conserved charge
follows immediately from (\ref{cch1}). The conserved current has the
same form as before: $J_a^{(1)}\equiv D_a\mu^{(0)}$, where $\mu^{(0)}$
is any constant, fiducial matrix.  However, because the spatial
topology of the reduced model is now trivial, the potential
$\mu^{(1)}$ is now defined simply by
\be 
 J_a^{(1)} = \epsilon_a^{\ b}\p_b \mu^{(1)}\ ,
\ee 
which determines $\mu^{(1)}$ up to an additive constant matrix,
\be 
\mu^{(1)} = -\int_0^r dr'\ J_0^{(1)}(r',t) 
              - \int_0^t dt'\ J_1^{(1)}(0,t')\ .
\ee

The second, fiducial current has the same form as $J_a^{(1)}$
\be 
   \kappa_a^{(2)} := D_a\mu^{(1)}\ . 
\ee
Taking its divergence, we find 
\bea
 \eta^{ab}\p_a\kappa^{(2)}_b &=& \epsilon^{ab}D_aJ_b^{(1)} 
= \epsilon^{ab}D_aD_b \mu^{(0)} \nn
&=& \left[F_{01},\mu^{(0)}\right] 
    = (1-r)\epsilon^{ab}\left[\p_aA_b, \mu^{(0)}\right] 
       + \left[A_0, \mu^{(0)}\right] \nn
&=& \epsilon^{ab}\p_a \left[(1-r)A_b, \mu^{(0)}\right] \nn
&=& \eta^{ab}\p_a\ \left[(1-r)A_c, \mu^{(0)}\right]\epsilon_b^{\ c}\ .
\eea
This allows us to identify the second conserved current
\be 
J_a^{(2)} = \kappa_a^{(2)} - (1-r)\epsilon_a^{\ b} 
\left[A_b,\mu^{(0)}\right] \ . 
\ee
Thus, the triviality of spatial topology simplifies the analysis. 
The corresponding conserved charge is 
\be 
Q^{(2)}= \int_0^\infty dr\ J_0^{(2)} 
= \int_0^\infty dr
\left\{ D_0\mu^{(1)} + (1-r)\left[A_1,\mu^{(0)}\right] \right\}\ .
\ee

As before, further conserved quantities can be generated by taking
Poisson brackets. Finally, in this case there is no global constraint
to take care of.


\section{Discussion}

Let us summarize the main results. In Sec. III, we first showed that
the standard, $SL(2,R)$ non-linear $\sigma$-model is embedded in the
principal chiral model in any space-time dimension. We then showed
that the symmetry reduced Einstein system is embedded in the $\sigma$
model in {\it three} space-time dimensions. Thus, there is a hierarchy
and results from the three-dimensional $\sigma$ and chiral models can
be taken over to the reduced Einstein system. This strategy is successful
in the one-polarization case, i.e., the case when the two Killing fields
are hypersurface-orthogonal.

In Sec IV we worked in two space-time dimensions and recast the
reduced Einstein model as a `time dependent' $\sigma$--model, or,
alternatively, `generalized' chiral model. For the standard chiral
model, there exist the so-called BIZZ procedure which enables one to
construct a hierarchy of conserved charges. In Sec. V, we showed that
the procedure can be appropriately modified to the generalized chiral
model of Sec. V to obtain the analogs of the first two BIZZ conserved
currents. Modifications were required because of two reasons: i) the
two-dimensional space-time topology is $S^1\times R$, rather than
$R^2$; and ii) the curvature of the generalized model does not vanish, 
but has a specific form. (To our knowledge, the complications that
arise due to non-trivial topology have not been discussed in the
literature, even in the standard BIZZ procedure.)  Since the currents
take values in the Lie-algebra of $SL(2,R)$, we obtain six conserved
quantities, $Q^{(1)}_i$ and $Q^{(2)}_i$, where the index $i$ refers to
the Lie-algebra. One can restrict this procedure to the
well-understood, one-polarization case. Even in this case, the second,
non-local set of charges $Q^{(2)}_i$ appears to be new, i.e., appears
to be a non-trivial combination of the known charges.

Returning to the general case, Poisson brackets between the
$Q^{(2)}_i$ lead to new conserved charges and one can continue the
process by taking Poisson brackets between the available charges. The
functional form of these charges exhibit increasing non-locality, each
step leading to an additional `nested integral'. The initial-value
problem for these equations is well-posed, and each charge appears to
probe a different aspect of the functional form of the initial
data. Therefore, it appears that the charges are all functionally
independent. However, we do not have a definitive proof of this
independence. For example, after a certain stage, the Poisson bracket
may just yield a c-number, i.e., a constant functional on the space of
solutions. In this case, the procedure would terminate. It may also
happen that at a certain stage the conserved charges cease to be
functionally differentiable, making it impossible to compute further
Poisson brackets. However, the non-local structure of charges is
reminiscent of the structure of the commutators in the Geroch
group. Is there a close relation between the two? If so, one would
have an elegant avenue to show that the charges are indeed
independent.

Finally, we should emphasize that, as matters stand, our results
have not lead to a comprehensive treatment of issues like exact
integrability or quantization in the full, two-polarization
case. Rather, they provide new windows to tackle these issues. In
particular, the techniques introduced in Secs. III and V should enable
one to analyze the reduced Einstein system along lines that are quite
different from the traditional ones.

{\it Note added}: After this work was completed and posted on the LANL
archives (gr/qc 9712053), we were informed of a paper by Romano and
Torre \cite{rt} which contains a result which is equivalent to that
presented in Sec. III.B. Their main interest is in the `issue of time'
and they work in a Hamiltonian framework with a phase space which is
enlarged to incorporate a `clock degree of freedom'. As a side-remark,
they point out that their Hamiltonian description can be derived from
a symmetry reduced three-dimensional harmonic map in the parametrized
field theory formalism.

\bigskip\bigskip

{\bf Acknowledgements:} This work was supported in part by the NSF
grant PHY 95-14240 and by the Eberly Research Funds of The
Pennsylvania State University. Work of V.H. was also supported in part
by the Natural Science and Engineering Research Council of Canada.


\end{document}